\documentclass{IEEEtran}
\usepackage{amsmath,amsfonts,amssymb,enumerate,cases}
\usepackage[final]{lscape,graphicx,epsfig}
\usepackage{cite}
\usepackage{subfigure}
\usepackage{caption}
\usepackage{amsthm}
\DeclareMathSizes{6}{6}{6}{6}

\DeclareMathOperator*{\Max}{max}

\newcommand{\NoiseSigma}{\sigma_{n}}

\newcommand{\onevect}{\mathbf{1}}

\newcommand{\La}{\mathbf{L}}
\newcommand{\D}{\mathbf{D}}

\newcommand{\A}{\mathbf{A}}

\newcommand{\smax}{\mathsf{smax}}

\newcommand{\Czero}{\mathbf{C}}
\newcommand{\CS}{\Czero}
\newcommand{\ao}{a^{*}}

\newcommand{\x}{\mathbf{x}}

\newcommand{\ybar}{\bar{y}}

\newcommand{\y}{\mathbf{y}}

\newcommand{\hatx}{\hat{x}}

\newtheorem{thm}{Theorem}

\allowdisplaybreaks

\title{Max Consensus in Sensor Networks: Non-linear Bounded Transmission and Additive Noise}

\author{ Sai Zhang, Cihan Tepedelenlio\u{g}lu, {\em Senior Member,
IEEE}, Mahesh K. Banavar, {\em Member,
IEEE}, and Andreas Spanias, {\em Fellow, IEEE} \thanks{The authors from Arizona State University are funded in part by the NSF award ECCS - 1307982 and the SenSIP Center, School of ECEE, Arizona State University. Banavar is funded in part by the NSF CRII award 1464222. (Email: \{szhang98, cihan, spanias\}@asu.edu, mbanavar@clarkson.edu). Parts of the material in this paper were presented at the 2013 Asilomar Conference on Signals, Systems and Computers \cite{wode_max_consensus}.}}

\date{}

\begin{document}

\maketitle

\begin{abstract}
A distributed consensus algorithm for estimating the maximum value of the initial measurements in a sensor network with communication noise is proposed. In the absence of communication noise, max estimation can be done by updating the state value with the largest received measurements in every iteration at each sensor. In the presence of communication noise, however, the maximum estimate will incorrectly drift and the estimate at each sensor will diverge. As a result, a soft-max approximation together with a non-linear consensus algorithm is introduced herein. 
A design parameter controls the trade-off between the soft-max error and convergence speed. An analysis of this trade-off gives a guideline towards how to choose the design parameter for the max estimate.
We also show that if some prior knowledge of the initial measurements is available, the consensus process can converge faster by using an optimal step size in the iterative algorithm. 
A shifted non-linear bounded transmit function is also introduced for faster convergence when sensor nodes have some prior knowledge of the initial measurements.
Simulation results corroborating the theory are also provided.
\end{abstract}


\section{Introduction} \label{sec: introduction}

In sensor networks with no fusion center, consensus is a process where all the sensors in the network achieve global agreement using only local transmissions.
There are many advantages of using decentralized wireless sensor networks, including robustness to link failures and scalability \cite{KarMoura2009, nonLinear2012Sivaraman}.


Max consensus algorithms in the absence of noise are considered in literature \cite{Consensus_switching_topo_and_time_delay_control, min_avr_consensus_EXP_3, max_consensus_analyze_2, max_consensus_update_1, Distributed_Averaging_and_Maximizing, max_consensus_asychronous_updates} with applications detailed in \cite{max_consensus_cognitive_radio, time_synchronization_app}. To the best of our knowledge, all past research on max consensus has assumed no communication noise. In the presence of noise, however, existing max consensus approaches will suffer from divergence.

In this paper, 
we design, for the first time in the literature, a fully distributed iterative algorithm to let nodes reach consensus on the maximum of the initial measurements in the presence of noise. 
The results provide guidelines on how to select a design parameter that balances convergence time with estimator accuracy.


\subsection{Literature Review on Existing Max Consensus Algorithms}
\label{sec: literature_review}

The distributed consensus problem has a long history and has attracted many researchers recently. It has broad applications in different areas \cite{synchronization_oscillator_networks, flocking_multi_agent_control, Survey_consensus_Multi_agent_Coordination, FunctionGM2009, nonLinear2012Sivaraman, KarMoura2009}.
A comprehensive review of the consensus literature is provided in \cite{literature_review_consensus_saber}. 
While average consensus is well studied in literature (e.g., \cite{KarMoura2009,Boyd2003,nonLinear2012Sivaraman}), estimating the average is not always the goal. In various applications, estimating the maximum measured value in the network is necessary \cite{FunctionGM2009}, \cite{max_consensus_update_1}. For example, spectrum sensing algorithms that use the OR-rule for cognitive radio applications can be implemented using max consensus \cite{max_consensus_cognitive_radio}. Also, max consensus can be used to estimate the maximum and minimum degrees of the network graph, which are useful in optimizing consensus algorithms \cite{Boyd2003}. In \cite{Consensus_switching_topo_and_time_delay_control}, it is also mentioned that max consensus and min consensus have a broad range of applications in distributed decision-making for multi-agent systems. In \cite{time_synchronization_app}, max consensus is used to compensate for clock drift and is used to time-synchronize wireless sensor network nodes.

To deal with the problem of finding a unique leader 
in a group of agents in a distributed way, a max consensus problem in a noise-free environment is proposed in \cite{Consensus_switching_topo_and_time_delay_control}, where each node in the network collects data from all of its neighbors and find the largest received data. At each iteration, after comparing its own state and the largest received data, each node updates its state with the max of the two values.

Max consensus algorithms using a similar approach as in \cite{Consensus_switching_topo_and_time_delay_control} are proposed in \cite{min_avr_consensus_EXP_3, max_consensus_analyze_2, max_consensus_update_1, Distributed_Averaging_and_Maximizing, max_consensus_asychronous_updates}. At each time step, every sensor in the network updates its state with the largest measurements it has recovered so far.
Reference \cite{max_consensus_update_1} considers both pairwise and broadcast communications, and analyzes the convergence time. 
A Max-plus algebra is used in \cite{max_consensus_analyze_2} to analyze the max consensus algorithm in a directed graph.
Time dependent graphs are considered in \cite{Distributed_Averaging_and_Maximizing}, where it is shown that strong connectivity is required for reaching max consensus.
A general class of algorithms which can be used for both average and min consensus algorithms is also mentioned in \cite{min_avr_consensus_EXP_3} by the selection of a design parameter. 


In \cite{Xconsensus_general_function}, the authors extend the work of the weighted power mean algorithm originally proposed by \cite{weightedPowerMean_optimal_distributed_consensus} and show that this class of algorithms can also be used to calculate the maximum of the initial measurements when the design parameter is chosen to be infinity. A similar max approximation algorithm is also mentioned in \cite{FunctionGM2009} to compute the maximum of the initial measurements in a centralized sensor network with a fusion center. 
Reference \cite{Xconsensus_general_function} also describes another distributed coordination algorithm for max consensus. 
None of these approaches addresses the issue of additive noise, or limited transmit power in their setup.

\subsection{Statement of Contributions} \label{sec: contributions}
In wireless sensor networks, information exchange between nodes usually involves communication noise and wireless fading channels, therefore it is practical to assume that communications between sensor nodes is noisy \cite{KarMoura2009, L_analysis_avg_consensus}.
To the best of our knowledge, all past research on max consensus has assumed no communication noise. 
In the presence of noise, if the max consensus algorithm at each node keeps the largest received data as in \cite{Consensus_switching_topo_and_time_delay_control, min_avr_consensus_EXP_3, max_consensus_analyze_2, max_consensus_update_1, Distributed_Averaging_and_Maximizing, max_consensus_asychronous_updates}, the estimated maximum value at each node will drift incorrectly at each iteration and states of nodes will not converge. The reason is that by using the $\max$ operator at each node, the positive noise samples will increase the maximum leading to divergence. 
Even with channel coding, 
codeword over the channel may cause divergence.
Some max approximation functions are considered in  \cite{min_avr_consensus_EXP_3, weightedPowerMean_optimal_distributed_consensus, FunctionGM2009, Xconsensus_general_function} for max consensus. However, the design parameters are chosen to be extreme values to get a small estimator error, without considering its effect on transmit power, or convergence time.
If the design parameter is set at extreme values, the the dynamic range of the state value will be extremely large, which results in large transmit power if linear consensus is used or slow convergence if non-linear consensus algorithm is used.

The contribution of this paper is in both design and analysis of a max consensus algorithm in wireless sensor networks in the presence of communication noise. 
Regarding design, the soft maximum, together with non-linear bounded transmissions is proposed. The soft maximum of a vector ${\x} := [x_1 \ldots x_N]$ is denoted as:
\begin{equation}
\label{eqn:soft_maximum} 
\smax(\x) = \frac{1}{\beta} \log \displaystyle\sum_{i=1}^{N} e^{\beta x_i},
\end{equation}
where $N$ is the number of elements in $\x$ and $\beta > 0$ is a design parameter. The soft-max of $\x$ approximates the maximum value of $\mathbf{x}$ for large $\beta$, and is equal to the maximum value of $\x$ as $\beta$ goes to infinity. The soft-max of $\x$ is always larger than the maximum value of $\x$ and the difference between the two values is less than $\beta^{-1}\log N$. Note that if $\beta < 0$, then \eqref{eqn:soft_maximum} approximates the minimum value of $\mathbf{x}$.

In the proposed max consensus algorithm, every sensor in the network evaluates a function $e^{\beta x_i}$ of its initial observation $x_i$ and a non-linear average consensus algorithm such as those in \cite{nonLinear2012Sivaraman} can be used with a judicious choice of $\beta$. 
The non-linear bounded transmission is used in the max consensus approach because an accurate max estimation result can be obtained only when the design parameter $\beta$ is chosen to be large. A large $\beta$ may result in a very large transmit power if the traditional linear consensus as in \cite{Boyd2003} is used. Since sensors typically have low transmit power, the bounded transmission assumption is a practical necessity. Note that since average consensus is used, the proposed max consensus algorithm will work even when the structure of the graph changes over time \cite{KarMoura2009}.


Regarding analysis, 
three sources of errors in the proposed max consensus algorithm are presented.
We show that the parameter of the soft-max function that makes the soft-max approximation accurate, also makes the convergence slow. The technical novelty in the analysis is the analytical study of this trade-off.
By bounding the error between the convergence result and the true max, we show that an approximation of the needed convergence time can be calculated. 
The last contribution of the paper is the introduction of the shifted non-linear bounded function, which makes the convergence faster. The analyses provide guidelines for nonlinear transmission design, and algorithm parameter settings to trade-off between estimation error and faster convergence.

\subsection{Organization}

The rest of this paper is organized as follows. In Section II, the system model is provided. We describe the problem statement in Section III, and prove that the sensors reach consensus with the proposed solution. The error analysis and convergence speed are also analyzed in Section III.
In Section IV, a shifted non-linear function is proposed for faster convergence. 
In Section V, simulation results for all proposed consensus algorithms are presented. Finally, concluding remarks are given in Section VI.

\section{System Model} \label{sec: system model}

\subsection{Graph Representation}
Consider an undirected connected graph $\mathbb{G}=(\mathbb{N}, \mathbb{E})$ containing a set of nodes $\mathbb{N}=\{1, \ldots, N\}$ and a set of edges $\mathbb{E}$.  The set of neighbors of node $i$ is denoted by $\mathbb{N}_{i}$, i.e., $\mathbb{N}_{i}=\{j|\{i,j\} \in \mathbb{E}\}$. Two nodes can communicate with each other only if they are neighbors. The number of neighbors of node $i$ is $d_{i}$. We use a degree matrix, $\D ={\rm diag} [d_{1},\;  d_{2}, \; \ldots, \; d_{N}]$, which is a diagonal matrix containing the degrees of each node. The connectivity structure of the graph is characterized by the adjacency matrix $\A=\{a_{ij}\}$ such that $a_{ij}=1$ if $\{i,j\} \in \mathbb{E}$ and $a_{ij}=0$ otherwise. The graph Laplacian of the network $\La$ is defined as $\La=\D - \A$. For a connected graph, the smallest eigenvalue of the graph Laplacian is always zero, i.e., $\lambda_1 (\La) =0$ and $\lambda_i (\La) > 0, i = 2, \cdots, N$. The zero eigenvalue $\lambda_1 (\La) =0$ corresponds to the eigenvector with all entries one, i.e. $\La \mathbf{1} = \mathbf{0}$. The performance of consensus algorithms often depends on $\lambda_2 (\La)$, which is also known as the algebraic connectivity \cite{OlfatiSaber2004}. Algebraic connectivity of simple and weighted graphs are discussed in \cite{graph_survey_2}, where several upper and lower bounds to $\lambda_2 (\La)$ are also given.


\subsection{Assumptions on Wireless Sensor Network Model}\label{sec: assumption_model}

We have the following assumptions on the system model: i) nodes in the distributed sensor network have their own initial measurements, and the nodes do not know if they have the maximum; ii) the communications in the network are synchronized, and at each iteration, nodes are broadcasting their state values to their neighbors; iii) communications between nodes is analog following \cite{literature_review_consensus_saber, Boyd2003, ultrafast_consensus_smallNetwork} and is subject to additive noise; and iv) each node updates its state based on the received data.

\section{Max Consensus Using the Soft-max} \label{sec: max consensus using soft-max}
\subsection{Problem Statement}\label{subsec:prob_state}

Consider a wireless sensor network with $N$ sensor nodes, each with a real-valued initial measurement, $x_i, i = 1,2, \cdots, N$. It is desired that the nodes reach consensus on the maximum value of the initial measurements, $x_{\rm max} := \Max_{i} x_i$, under the assumption that the sensors have a single state that they update based on local received measurements. Max consensus in the absence of noise is straight forward: the nodes update their states with the largest received measurement thus far in each iteration.
Consider the following algorithm at each node:
\small
\begin{align}
\label{eq: naive_max_consensus_iteration_1}
& x_{i}(t \!+\! 1) \!=\! \max \left\lbrace  x_{i}(t), \max_{j \in \mathbb{N}_{i}} x_{j}(t)   \right\rbrace, \; \hat{x}_{\max, i}(t \!+\! 1) \!=\! x_{i}(t \!+\! 1).
\end{align}
\normalsize
However, in the presence of noise, such algorithms will diverge due to positive noise samples, which can be shown in Figure~\ref{Fig02} in Section~\ref{sec: simualtions}.
An intuitive explanation is that and any positive noise sample will always make the maximum larger if the $\max$ operator is used in the max consensus algorithm.




\begin{figure}[!t]
\centering
    \hspace{-0.5cm}
    \includegraphics[height=4.15cm,width=8.05cm]{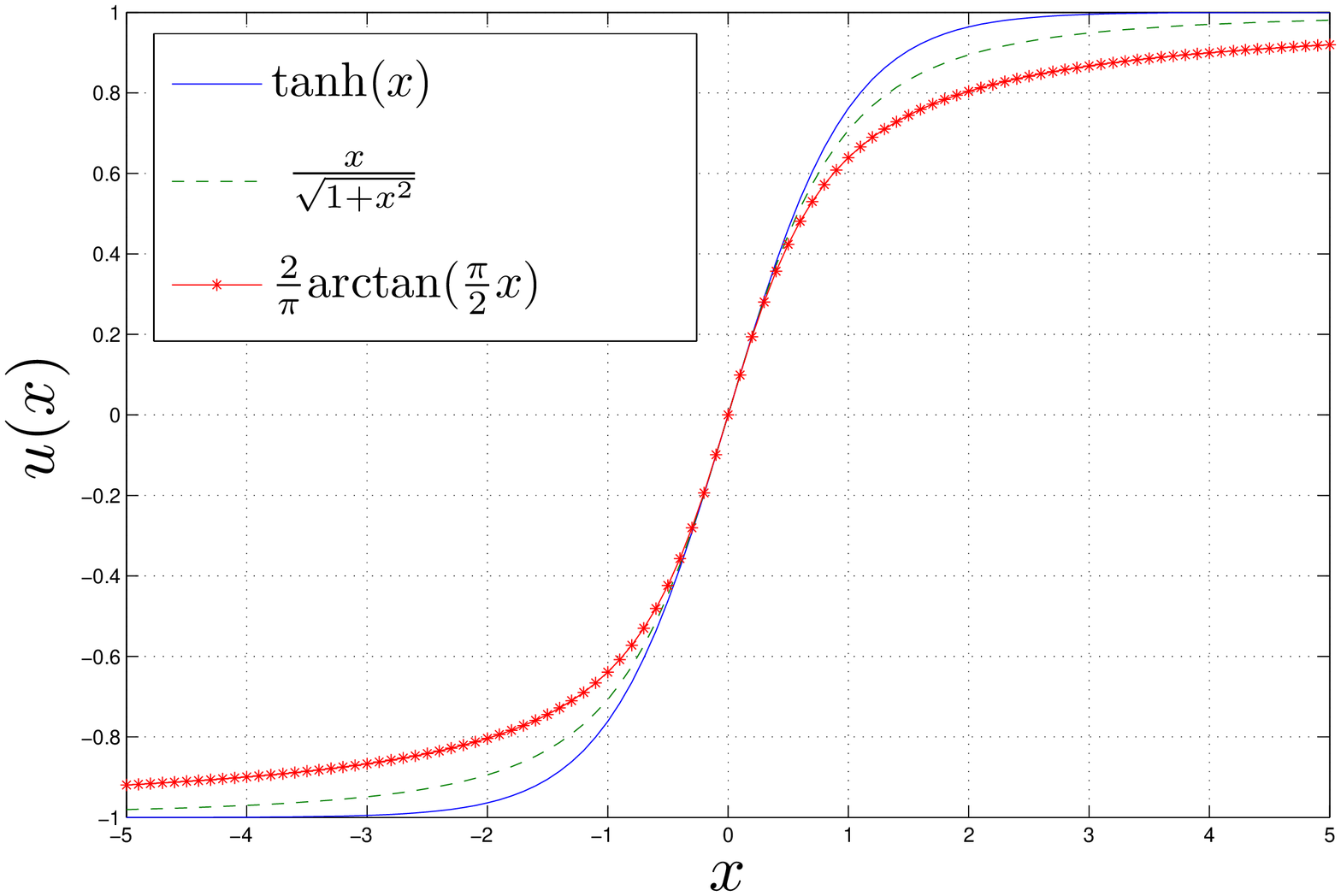}
\caption{Bounded Transmission Functions.}
\label{Fig_sigmoid} 
\end{figure}

Average consensus is well studied in literature. Existing average consensus algorithms converge to the sample mean of the initial measurements. As a result, the soft-max can be used to calculate the maximum. To relate the soft-max to the sample mean of $\{ e^{\beta x_i} \}$, we have,
\small
\begin{equation}
\label{eq:sample_avg}
\ybar = \frac{1}{N} \displaystyle\sum_{i=1}^{N} e^{\beta x_i} = \frac{1}{N} \displaystyle\sum_{i=1}^{N} y_i (0),
\end{equation}
\normalsize
where $\ybar$ is the sample mean of the mapped initial measurements and $y_i (0) := e^{\beta x_i}$. The quantity $\ybar$ is computed using an iterative distributed algorithm, in which each sensor communicates only with its neighbors. If the states of all the sensor nodes converge to  $\ybar$, then the network is said to have reached consensus on the sample average of the mapped initial measurements. The relation between $\ybar$ and the soft-max value is given by
\small
\begin{equation}\label{eqn:soft_maximum2} 
\smax( \mathbf{x} ) = \frac{1}{\beta} \log \displaystyle\sum_{i=1}^{N} e^{\beta x_i}\ = \frac{1}{\beta} (  \log N + \log \ybar  ).
\end{equation}
\normalsize

The average consensus algorithms like in \cite{KarMoura2009, nonLinear2012Sivaraman} can be used to achieve consensus in the sensor network. Sensors may adopt either a digital or analog method for transmitting their information to their neighbors. One such method is the linear amplify-and-forward (AF) scheme in which sensors transmit scaled versions of their measurements to their neighbors where the iterative algorithm may be chosen as the linear consensus algorithm of \cite{KarMoura2009}. However, using the AF technique is not a viable option for consensus on the soft-max. The reason is that accurate approximation of the max value using the soft-max method requires the parameter $\beta$ to be large, which can result in a large dynamic range of the mapped initial measurements and large transmit power. Moreover, using a linear transmit amplifier is power-inefficient. As a result, a non-linear consensus (NLC) algorithm can be implemented \cite{nonLinear2012Sivaraman}. The consensus on the soft maximum is achieved by letting each sensor map its state value at time $t$ through a bounded function $h(\cdot)$ before transmission to ensure bounded transmit power. To describe the communications between nodes, we use the standard Gaussian MAC so that each node receives a noisy version of the superposition of the transmitted signal from its neighbors. This is because the step sizes are the same across different network links and there is no need to recover the transmitted data separately. Consider the following algorithm with additive noise at the receiver:
\small
\begin{equation}
\label{eq:nld_ch_noise}
y_{i}(t  \!+\!  1) = y_{i}(t) - \alpha(t) \Big[ d_i h(y_{i}(t)) - \displaystyle\sum_{j \in \mathbb{N}_{i} } h(y_{ij}(t)) + n_{i}(t) \Big],
\end{equation}
\normalsize
where $i=1, 2, \ldots, N$, and $t=0,1,2,\ldots$, is the time index. The value $y_{i}(t+1)$ is the state update of node $i$ at time $t+1$, $y_{ij} (t)$ is the state value of the $j^{th}$ neighbor of node $i$ at time $t$, and $n_{i}(t)$ is the additive noise at node $i$, which is assumed to be independent across time and space with zero mean and variance $\NoiseSigma^2$. $\alpha(t)$ is a positive step size which satisfies $\sum_{t=0}^{\infty} \alpha^{2}(t) < \infty$ and $\sum_{t=0}^{\infty} \alpha(t) = \infty$. The node $j$ transmits its information $y_{ij}(t)$ by mapping it through the non-linear function $h(\cdot)$ to constrain the transmitted power. We assume that
\small
\begin{equation}
\label{eq: def_hx}
h(x) = \sqrt{\gamma} \;u(\omega x) ,
\end{equation}
\normalsize
where $u(x)$ is a normalized non-linear bounded function as in Figure~\ref{Fig_sigmoid} and we make the following assumption on $u(x)$:\\
\textbf{Assumptions}\\
\textbf{(A1):} $u(0) = 0. \; u(x) = - u(-x)$.\\
\textbf{(A2):} $\max\left( u(x) \right) = 1$.\\
\textbf{(A3):} The function $u(\cdot)$ is differentiable and invertible, $u^{\prime}(0) = 1$ and $0 < \frac{d u(x)}{d x} \leq 1$.\\
Parameters $\gamma$ and $\omega$ are constants used to control the shape of $h(\cdot)$. Note that invertibility of $h(\cdot)$ is needed for convergence, however there is no need to apply the inverse of $h(\cdot)$ in equation \eqref{eq:nld_ch_noise}.

Node $i$ receives a noisy version of the superposition $\sum_{j \in \mathbb{N}_{i}}h(y_{ij}(t))$. The recursion in equation \eqref{eq:nld_ch_noise} can be expressed in vector form as,
\small
\begin{equation}
\label{eq:nld_ch_noise_vectorForm}
\mathbf{y}(t+1) = \mathbf{y}(t) - \alpha(t)[\La h(\mathbf{y}(t)) + \mathbf{n}(t)],
\end{equation}
\normalsize
where $\mathbf{y}(t) = [y_{1}(t)\;  y_{2}(t) \; \cdots \; y_{N}(t)]^{\rm T}$ and $h(\mathbf{y}(t)) = [h(y_{1}(t)) \; h(y_{2}(t)) \; \cdots \; h(y_{N}(t))]^{\rm T}$. $\La$ is the Laplacian matrix of the graph and $\mathbf{n}(t)$ is the vector containing the additive reception noise at nodes. Since the noise is i.i.d. with variance $\NoiseSigma^2$, the covariance of $\mathbf{n}(t)$ is $\NoiseSigma^2 \mathbf{I}$.
Since \eqref{eq:nld_ch_noise} converges to a value that approximates \eqref{eq:sample_avg}, the consensus estimate of the maximum at node $i$ can be written using \eqref{eqn:soft_maximum2} as
\small
\begin{equation}\label{eq:xmax_estimate}
{\hat{x}}_{{\rm max}_i} (t^{*}) = \frac{1}{\beta} \left(  \log N + \log y_{i}(t^{*})  \right),
\end{equation}
\normalsize
where $t^{*}$ is the iteration at which the algorithm is stopped.

\subsection{Proof of Convergence}\label{subsec:prove of the convergence}

Since the non-linear average consensus approach is used in the max consensus algorithm, the convergence proof will follow the proof in \cite{nonLinear2012Sivaraman} which uses a discrete time Markov process approach \cite{stochastic_approximaton_recursive_estimation} (also see Theorem 5 in \cite{nonLinear2012Sivaraman}). Therefore, there exists a finite real random variable $\theta^{*}$ such that, 
\small
\begin{equation}\label{eq:state converge}
\Pr \left[ \lim_{t\rightarrow \infty} \y \left(t\right) = \theta^{*} \onevect \right] = 1,
\end{equation}
\normalsize
where $\onevect$ is a column vector with all ones. Equation \eqref{eq:state converge} shows that the convergence is reached when $t \rightarrow \infty$. 
The following theorem characterizes the random variable $\theta^{*}$.

\newtheorem{thm1}{thm1}
\begin{thm} 
\label{thm1}
$\theta^{*}$  in \eqref{eq:state converge} is an unbiased estimate of $\ybar$, $\mathrm{E}[\theta^{*}] = \bar{y}$. Its mean square error $\xi_{N} = \mathrm{E} [ ( \theta^{*} - \overline{y} )^{2} ]$, and is finite which can be expressed as,
\small
\begin{equation}
\label{eq:MSE error}
\xi_{N} = \frac{\NoiseSigma^2}{N} \sum_{t= 0}^{\infty} \alpha^{2}(t).
\end{equation}
\normalsize
\end{thm}

Proof: The proof is a straightforward adaptation of Theorem 3 in \cite{nonLinear2012Sivaraman}.

The nodes in the sensor network reach consensus on the random variable $\theta^{*}$ which is an unbiased estimator of the average of the mapped initial measurements,  $\mathrm{E}[\theta^{*}] = \ybar$. Then the soft-max of the initial measurements can be obtained using equation \eqref{eqn:soft_maximum2}.

In the following subsection, we will analyze the performance of the system by considering the three sources of error between the max estimate at each node $\hat{x}_{{\rm max}_i} (t^{*})$ in \eqref{eq:xmax_estimate}, and the true maximum of the initial measurements $x_{\rm max}$.

\subsection{Error Analysis}\label{subsec:error analysis}

Let $\theta_0$ be a realization of $\theta^{*}$. From \eqref{eq:state converge} we have that the states of nodes in the sensor network are converging to $\theta_0$ as $t \rightarrow \infty$. However, in practice, we need to stop the algorithm at a finite iteration time $t^{*}$. There are three sources of error between the true maximum $x_{\rm max}$ and ${\hat{x}}_{{\rm max}_i} (t^{*})$ in \eqref{eq:xmax_estimate}: i) $\left( \smax( \mathbf{x} ) - x_{\rm max} \right) = \frac{1}{\beta} \left( \log N + \log \bar{y} \right) - x_{\rm max}$, due to the fact that soft-max approximation will always be larger than the true max; ii) $\left( \theta_0 - \bar{y} \right)$ caused by communication noise and iii) $\left( y_{i}(t^{*}) - \theta_0 \right)$ cause by finite number of iterations.

The size of these errors depend on $\beta, N, t^{*}$ and $\NoiseSigma^{2}$. In the following subsection, we are going to characterize and analyze these errors.

\subsubsection{Soft-max error}\label{bound for softmax}

This is a deterministic error which depends on $\beta, N$, and the value of $\mathbf{x}$. We have:
\small
\begin{equation}\label{eq:bound of softmax}
x_{\rm max} \leq \smax (\x) \leq x_{\rm max}  + \dfrac{1}{\beta} \log N,
\end{equation}
\normalsize
Both inequalities are clearly tight for large $\beta$.

\subsubsection{MSE of the algorithm}\label{mse of non-linear transmission}

The second term $\left( \theta_0 - \bar{y} \right)$ is due to the presence of communication noise: the state of the sensors does not converge to the sample mean of the mapped initial measurements, instead it converges to a random variable $\theta^{*}$ whose expectation is the sample mean of the mapped initial measurements, $\ybar$ from \eqref{eq:sample_avg}. This occurs also in linear average consensus in the presence of noise. The mean square error of $\theta^{*}$ is defined as $\xi_{N} = E [ ( \theta^{*} - \overline{y} )^{2} ]$ and is characterized as \eqref{eq:MSE error} in Theorem \ref{thm1}.
From \eqref{eq:MSE error}, we see that the mean square error is finite and is small when $\sum_{t=0}^{\infty} \alpha^{2}(t)$ or $\NoiseSigma^{2}$ small. 

\subsubsection{Convergence speed}\label{asymptotic covariance matrix}

The third cause of error is due to a finite number of iterations: even though $\lim_{t \rightarrow \infty} y(t) = \theta_0$, $y(t^{*}) \neq \theta_0$. However, with a judicious choice of non-linear function $h(\cdot)$ and step size $\alpha(t)$, one can reduce the convergence time. From now on, we will assume that $\alpha(t) = \frac{a}{t+1}, a>0$, which satisfies $\sum_{t=0}^{\infty} \alpha^{2}(t) \leq \infty$, $\sum_{t=0}^{\infty} \alpha(t) = \infty$. The convergence speed is analyzed by establishing $\sqrt{t}(\mathbf{y}(t) - \theta_0 \onevect)$ is asymptotically normal with zero mean and some covariance matrix $\mathbf{C}$. The next theorem further quantifies the convergence speed.

\newtheorem{thm2}{thm2}
\begin{thm}
\label{thm2}
Let $2a \lambda_2(\La) h^{\prime}(\theta_0) > 1$ so that the matrix $\left[ a h^{\prime}(\theta_0) \mathbf{B} + \mathbf{I}/2\right]$ is stable (every eigenvalue of the square matrix has strictly negative real part) and $\mathbf{I}$ is the identity matrix, and $\mathbf{B}$ is a diagonal matrix containing all the non-zero eigenvalues of $-\La$. Define $\mathbf{U} = [ N^{-1/2} \mathbf{1} \; \mathbf{\Phi}]$ which is a unitary matrix whose columns are the eigenvectors of $\La$. Let $\left[ \tilde{n}(t) \; \tilde{\mathbf{n}}(t) \right] = N^{-1} \mathbf{U}^{\rm T} \mathbf{n}(t)$ and $\mathbf{C}_{\tilde{\mathbf{n}}} = {\rm E}[\tilde{\mathbf{n}} \tilde{\mathbf{n}}^{\rm T}]$ is a diagonal matrix, $\mathbf{C}_{\tilde{\mathbf{n}}} \in \mathbb{R}^{(N-1) \times (N-1)}$. Then as $t \rightarrow \infty$,
\small
\begin{equation}
\label{eq: show_distribution_ASMY}
\sqrt{t}(\mathbf{y}(t) - \theta_0 \mathbf{1}) \sim \mathcal{N}(\mathbf{0}, \mathbf{C}),
\end{equation}
\normalsize
where the asymptotic covariance matrix
$\mathbf{C} = N^{-1}a^2 \NoiseSigma^2 \mathbf{1} \mathbf{1}^{\rm T} + N^{-1} \mathbf{\Phi} \mathbf{S}^{\theta_0} \mathbf{\Phi}^{\rm T}$,\\ and
$\mathbf{S}^{\theta_0} = a^2 \int_{0}^{\infty} e^{\left( ah^{\prime}(\theta_0)\mathbf{B} + \mathbf{I}/2 \right) t} \mathbf{C}_{\tilde{\mathbf{n}}} e^{\left( ah^{\prime}(\theta_0)\mathbf{B} + \mathbf{I}/2 \right) t}$.
\end{thm}

The proof is the same as given in Theorem 5 in \cite{nonLinear2012Sivaraman}.

The convergence speed is quantified by $\| \mathbf{C} \|$, which is defined to be the largest eigenvalue of the covariance matrix. We show in Appendix that the $l_{2}$ norm of the covariance matrix can be expressed as
\small
\begin{align}
\label{eq: normCovariance_largestEigenvalue_result}
\| \mathbf{C} \| & = \max_{\|\x \| \leq 1} \x^{\rm T} \mathbf{C} \x \notag \\
& = \max \left\lbrace a^2 \NoiseSigma^2, \; \frac{1}{N}\frac{ a^2 \NoiseSigma^2}{2a h^{\prime}(\theta_{0})\lambda_{2}(
\La) - 1} \right\rbrace.
\end{align}
\normalsize
This norm, $\| \mathbf{C} \|$, can be optimized with respect to $a$, and the value that minimizes $\| \mathbf{C} \|$ is $a^{*} = (N+1)/[2N\lambda_{2}(\La)h^{'}(\theta_{0})]$. The optimal value for the $l_{2}$ norm of the covariance matrix denoted as $\| \mathbf{C}^{*} \|$ can be represented as
\small
\begin{align}
\label{eq:asymp_covariance_value}
 \| \mathbf{C}^{*} \| & = \left (\frac{N+1}{2 N} \right)^2 \left (\frac{\sigma^2_{n} }{\lambda^2_2(\La)} \right )  \left ( \frac{1}{h^{'}(\theta_0)} \right )^2 \notag \\
 & = \left (\frac{N+1}{2 N} \right)^2 \left (\frac{\sigma^2_{n} }{\lambda^2_2(\La) \gamma} \right )  \left ( \frac{1}{\omega u^{'}(\omega \theta_0)} \right )^2,
\end{align}
\normalsize
which is proved in Appendix. The interpretation is that convergence is slower when $ \| \mathbf{C}^{*} \|$ is larger.

It is clear form equation \eqref{eq:asymp_covariance_value} that convergence will be fast if $\lambda_2(\La)$ large, which implies faster convergence in a more connected graph. Also the value of $\|\mathbf{C}^{*}\|$ decreases as $h^{'}(\theta_0)$ increases, which shows that the convergence speed depends on the non-linear function and the convergence point.

By observing the three sources of error mentioned above, we find there is a trade-off between the convergence speed and the soft-max error. To see this, recall that the convergence speed is quantified by $\| \mathbf{C}^{*} \|$. From the analysis of sources of error, choosing a larger $\beta$ would reduce the deterministic bias caused by the soft-max mapping \eqref{eq:bound of softmax}, but degrades the variance term in \eqref{eq:asymp_covariance_value}. The reason is that $h(\cdot)$ is chosen to be an odd bounded transmission function as in Figure~\ref{Fig_sigmoid} with a zero-crossing and steepest slope at the origin, with $h^{\prime}(x)$ decreasing for $x \geq 0$. Since $\theta_0 \geq 0$, $h^{'}(\theta_0)$ will be small when $\theta_0$ gets larger which increases the value of $ \| \CS^{*} \|$ and makes the convergence slower. The convergence point $\theta_0$ will be large when $\beta$ is chosen large since $\theta_{0} \approx \frac{1}{N} \sum_i e^{\beta x_i}$. Therefore a trade-off between the convergence speed and the soft-max error exists: a more accurate soft-max can be obtained by choosing a large $\beta$, but this degrades the convergence speed.

\subsection{Bound on Convergence Time}\label{Bound on Convergence Time}

The convergence speed of the max-consensus algorithm is quantified by the asymptotic covariance matrix. If some prior knowledge about the distribution of the initial measurements is known, the step size can be set based on the expression of $\ao = (N+1)/[2N\lambda_{2}(\La)h^{'}(\theta_{0})]$ and $\alpha(t) = \ao /(t+1)$ to make the convergence fast. In this section, we assume that the step size is set to be $\ao$ as mentioned. The trade-off controlled by $\beta$ balances soft-max error and convergence speed. How much time $t^{*}$ is needed for the nodes to reach consensus is always an important problem. In this subsection, we will show that by upper bounding the three sources of error in Section~\ref{subsec:error analysis}, an approximation on the iteration time for reaching consensus can be calculated.

The estimate of the max at iteration time $t^{*}$ is expressed as \eqref{eq:xmax_estimate}, where $y_{i}(t^{*})$ is the state at node $i$ at time $t^{*}$. Of the three errors in Section~\ref{subsec:error analysis}, 
note that the error $(\theta_0 - \bar{y})$ can be ignored when the noise variance $\NoiseSigma^2$ is small, or can be reduced by running the consensus several times and taking the average of the results. In the following, we ignore the error $(\theta_0 - \bar{y})$ and calculate the iteration time by bounding the soft-max error denoted by $\varepsilon_2$ and error caused by a finite stopping time $t^{*}$ denoted by $\varepsilon_1$.
When $a = \ao$, the norm of the asymptotic covariance matrix of $(\mathbf{y}(t^{*}) - \theta_0 \onevect)$ can be bounded by $\varepsilon_1$ if
\small
\begin{equation}
\label{eq: bound_asympotic_C_1}
\frac{\| \CS^{*} \|}{t^{*}} \leq \varepsilon_1 \Rightarrow t^{*} \geq \frac{\| \CS^{*} \|}{\varepsilon_1}.
\end{equation}
\normalsize
The soft-max error is bounded by bounding the upper bound in equation \eqref{eq:bound of softmax}, which can be expressed as:
\small
\begin{equation}
\label{eq: bound_soft_max_2}
\frac{\log N}{\beta} \leq \varepsilon_2 \Rightarrow \beta \geq \frac{\log N}{\varepsilon_2}.
\end{equation}
\normalsize
By substituting \eqref{eq: bound_soft_max_2} into $\| \CS^{*} \|$, a lower bound of the iteration time needed for reaching consensus can be calculated using \eqref{eq: bound_asympotic_C_1}:
\small
\begin{align}
\label{eq: iteration_time_lower_bound_calculated}
&t^{*} \geq \frac{\| \CS^{*} \|}{\varepsilon_1} = \left( \frac{N+1}{2 N} \right)^2 \left( \frac{\sigma^2_{n} }{\lambda^2_2(\La)} \right) \frac{ \left( \frac{1}{h^{'}(\theta_0)} \right)^2}{\varepsilon_1} \notag \\
&= \left( \frac{N+1}{2 N} \right)^2 \left( \frac{\sigma^2_{n} }{\lambda^2_2(\La)} \right) \frac{ \left( \frac{1}{h^{'}(\frac{1}{N} \sum_i e^{\beta x_i})} \right)^2}{\varepsilon_1} \notag \\
&\geq \left(\! \frac{N\!+\!1}{2 N} \!\right)^2 \left(\! \frac{\sigma^2_{n} }{\lambda^2_2(\La)} \!\right) \left(\! \frac{1}{\varepsilon_1} \!\right) \left( \frac{1}{h^{'}\left( \frac{1}{N} \sum_i e^{\frac{\log N}{\varepsilon_2}x_i} \right) }\right) ^2.
\end{align} 
\normalsize
The last inequality holds because of \eqref{eq: bound_soft_max_2} and using that $h^{\prime}(x)$ is decreasing function when $x > 0$. We now study how the final lower bound depends on the convergence error, $\varepsilon_1$, and soft-max error, $\varepsilon_2$. It is clear that the bound is inversely related to $\varepsilon_1$. 
How $\varepsilon_2$ affects the bound depends on the choice of $h(\cdot)$.

In the following, we provide two examples of $h(\cdot)$ and show that how equation \eqref{eq: iteration_time_lower_bound_calculated} is affected by $\varepsilon_2$.
We will consider two cases. In the first case, $h(x)$ converges to its maximum value polynomially fast and in the second case it converges exponentially.
First consider the polynomial case and let $h(x) \approx \sqrt{\gamma} \left( 1 - \frac{1}{x^{p} + 1} \right)$ for large $x > 0$ for $p > 0$. Note that $p$ controls the value of $\omega$ in the definition of $h(x)$ in \eqref{eq: def_hx}. Then, 
\small
\begin{align}
\label{eq: ppolynomial_calculation_process_how_decrease_1}
&h^{\prime}\left( \frac{1}{N} \sum_i e^{\frac{\log N}{\varepsilon_2} x_i} \right) \approx h^{\prime}\left( \frac{1}{N} e^{\frac{\log N}{\varepsilon_2} x_{\rm max}} \right) \\
\label{eq: ppolynomial_calculation_process_how_decrease_2}
& = -p \sqrt{\gamma} \left(\log N\right) x_{\rm max} \frac{\varepsilon_{2}^{-2} N^{p(\frac{x_{\rm max}}{\varepsilon_2} - 1)} }{\left( N^{p(\frac{x_{\rm max}}{\varepsilon_2} - 1)} + 1 \right)^2} \\
\label{eq: ppolynomial_calculation_process_how_decrease_3}
& \approx - \frac{p \sqrt{\gamma} \left(\log N\right) x_{\rm max}}{\varepsilon_{2}^{2} N^{p(\frac{x_{\rm max}}{\varepsilon_2} - 1)}}. 
\end{align}
\normalsize
Equation \eqref{eq: ppolynomial_calculation_process_how_decrease_1} holds since when $\varepsilon_2$ is small, the term $x_i = x_{\rm max}$ dominates. Equation \eqref{eq: ppolynomial_calculation_process_how_decrease_3} shows how $\varepsilon_2$ affects the convergence time when $h(x) \approx \sqrt{\gamma} \left( 1 - \frac{1}{x^{p} + 1} \right)$ for large $x > 0$ for $p>0$. 
The asymptotically optimal $p$ that minimizes \eqref{eq: iteration_time_lower_bound_calculated} is $p^{*} = 1/\left( (\log N) \left( \frac{x_{\rm max}}{\varepsilon_2} - 1\right) \right)$, and the lower bound of the iteration time needed can be calculated as,
\small
\begin{equation}
\label{eq: Optimal_polynomial_case1}
t^{*} \geq \left(\! \frac{N\!+\!1}{2 N} \!\right)^2 \left(\! \frac{\sigma^2_{n} }{\lambda^2_2(\La)} \!\right) \left(\! \frac{1}{\varepsilon_1} \!\right) \left(\! \frac{\varepsilon_{2}^{4} e^2 \left( \frac{x_{\rm max}}{\varepsilon_2} - 1 \right)^2 }{ \gamma x^{2}_{\rm max}} \!\right).
\end{equation}
\normalsize

On the other hand, if $h(x)$ converges to its final value, $\sqrt{\gamma}$, exponentially fast, we have $h(x) \approx \sqrt{\gamma} \left( 1 - e^{-q x} \right)$ for large $x > 0$, with $q > 0$ which controls the value of $\omega$ in the definition of $h(x)$ in \eqref{eq: def_hx}. Then,
\small
\begin{align}
\label{eq: exponential_calculation_process_how_decrease_1}
&h^{\prime}\left( \frac{1}{N} \sum_i e^{\frac{\log N}{\varepsilon_2} x_i} \right) \approx h^{\prime}\left( \frac{1}{N} e^{\frac{\log N}{\varepsilon_2} x_{\rm max}} \right) \\
\label{eq: exponential_calculation_process_how_decrease_2}
& = -q \sqrt{\gamma} \frac{\log N}{N} x_{\rm max} \left( \varepsilon_2^{-2} N^{\frac{x_{\rm max}}{\varepsilon_2}} \right) \left( e^{-q N^{\left( \frac{x_{\rm max}}{\varepsilon_2} - 1\right)} } \right) \\
\label{eq: exponential_calculation_process_how_decrease_3}
& = - \frac{ q \sqrt{\gamma}\left(\log N\right) x_{\rm max} N^{\left( \frac{x_{\rm max}}{\varepsilon_2} - 1\right) }  }{ \varepsilon_2 e^{q N^{\left( \frac{x_{\rm max}}{\varepsilon_2} - 1 \right)} } }.
\end{align}
\normalsize
The asymptotically optimal $q$ that minimizes \eqref{eq: iteration_time_lower_bound_calculated} is $q^{*} = N^{\left(   1 - \frac{x_{\rm max}}{\varepsilon_2}\right)}$, and the lower bound of the iteration time needed can be calculated as,
\small
\begin{equation}
\label{eq: Optimal_exponential_case2}
t^{*} \geq \left( \frac{N+1}{2 N} \right)^2 \left( \frac{\sigma^2_{n} }{\lambda^2_2(\La)} \right) \left( \frac{1}{\varepsilon_1} \right) \left( \frac{\varepsilon_{2}^{4} e^2 }{ \gamma x^{2}_{\rm max}  \left( \log N\right)^2 } \right).
\end{equation}
\normalsize

When choosing the non-linear bounded function as mentioned above, we have the following observations based on equation \eqref{eq: iteration_time_lower_bound_calculated}, \eqref{eq: ppolynomial_calculation_process_how_decrease_3} and \eqref{eq: exponential_calculation_process_how_decrease_3}: (i) the required convergence time will be longer when the error requirements $\varepsilon_1$ and $\varepsilon_2$ are smaller; (ii) the soft-max error term $\varepsilon_2$ dominates the convergence time in equation \eqref{eq: iteration_time_lower_bound_calculated} in both examples; (iii) the required convergence time will be longer when $x_{\rm max}$ or the system size $N$ is larger; (iv) by comparing equation \eqref{eq: Optimal_polynomial_case1} and \eqref{eq: Optimal_exponential_case2}, the convergence will be faster when choosing $h(\cdot)$ that converges to its maximum value exponentially fast is appropriate if
\small
\begin{equation}
\left( \frac{x_{\rm max}}{\varepsilon_2} - 1 \right)^2 > \left( \log N\right)^{-2},
\end{equation}
\normalsize
and $h(\cdot)$ that converges to its maximum value polynomially should be chosen otherwise.

Finally, note that estimating the minimum value of the local measurements is also sometimes necessary. The min-consensus can be achieved using the similar initial mapping but choosing $\beta < 0$.

\section{Shifted Non-linear Bounded Function Used in Max Consensus} \label{sec: shifted h function}

An accurate max estimation using the soft-max approach requires the design parameter $\beta$ to be large. As a result, the exponential function used for initial measurements mapping expands the dynamic range of the initial measurements and the convergence speed is slow. We now give a modified non-linear distributed average consensus method using a shifted non-linear bounded function that can make the convergence process faster if some prior knowledge of the initial measurements is available.

The method is based on the fact that the convergence is faster when the value of $h^{'}(\theta_{0})$ is large from equation \eqref{eq:asymp_covariance_value}. $h(\cdot)$ in the iterative algorithm is replaced by a shifted non-linear function $g(\cdot)$ defined as $g(x) = h(x - T)$, where $T$ is a shift constant.
In this case, the optimal step size $a^{*}_{\rm s} = \left( \frac{N+1}{2N} \right) \left( \frac{1}{\lambda_{2}(\La) g^{\prime}(\theta_{0})} \right) = \left( \frac{N+1}{2N} \right) \left( \frac{1}{\lambda_{2}(\La) h^{\prime}(\theta_{0}- T)} \right)$. The convergence speed is quantified by the norm of the asymptotic covariance matrix $\| \mathbf{C}_{\rm s}^{*}\|$ and can be expressed as,
\small
\begin{equation}
\label{eq: convergence_speed_NLCwithSHIFT}
\| \mathbf{C}_{\rm s}^{*}\| = \left (\frac{N+1}{2 N} \right)^2 \left (\frac{\sigma^2_{n} }{\lambda^2_2(\La)} \right )  \left ( \frac{1}{h^{'}(\theta_0 - T)} \right )^2.
\end{equation}
\normalsize
From \eqref{eq: convergence_speed_NLCwithSHIFT}, convergence will be faster when $h^{'}(\theta_0 - T)$ is larger. $h^{'}(\theta_0 - T)$ reaches its largest value when $T = \theta_{0}$ if $h(\cdot)$ is chosen as a sigmoid function with steepest slope at origin.
Note that $\theta_0$ is unknown in practice, but one can use  prior information on the initial measurements to choose $T$.
If the distribution of the initial measurements is known at the sensor nodes, a reasonable choice of $T$ is to choose it as the expected value of the mapped initial measurements: $T = E[e^{\beta x_i}] \approx \frac{1}{N} \sum_{i=1}^{N} e^{\beta x_i}$. 


\section{Simulations}
\label{sec: simualtions}

In this section, simulation results for max consensus algorithms are presented. Different $\beta$ values are used to trade-off between convergence speed and error between the proposed approach and the true max of the initial measurements.

\begin{figure}
	\centering
    \includegraphics[height=4.85cm,width=6.85cm]{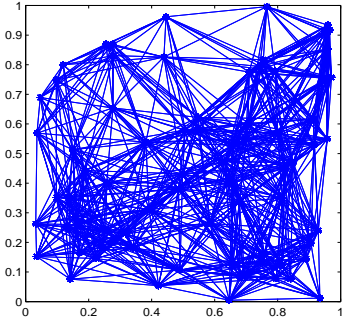}
\caption{Graph Representation of the Sensor Network, $N = 75$.}
\label{Fig01} 
\end{figure}

\begin{figure}
	\centering
    \includegraphics[height=4.85cm,width=7.45cm]{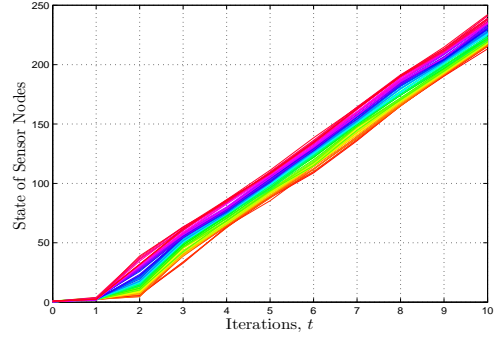}
\caption{Entries of traditional max consensus result versus iterations $t$ (Keep the largest measurement at each iteration).}
\label{Fig02} 
\end{figure}

\begin{figure}
\centering
\subfigure[Entries of the consensus soft max result versus iterations $t$, $\beta$ = 5, $\omega$ = 0.015, $h(x)$ = $\sqrt{\gamma} \tanh(\omega x)$, $\alpha(t)\! =\! \frac{4.4473}{t+1}$, $\ao \approx 4.4473$.] 
{
    \label{Fig22}
    \includegraphics[height=4.85cm,width=7.45cm]{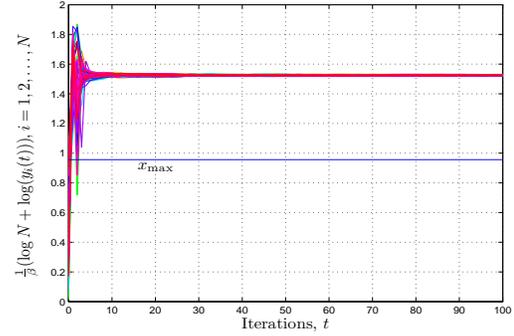}
}
\centering
\subfigure[Entries of the consensus soft max result versus iterations $t$, $\beta$ = 7, $\omega$ = 0.015, $h(x)$ = $\sqrt{\gamma} \tanh(\omega x)$, $\alpha(t)\! =\! \frac{61.7513}{t+1}$, $\ao \approx 61.7513$.] 
{
    \label{Fig24}
    \includegraphics[height=4.85cm,width=7.45cm]{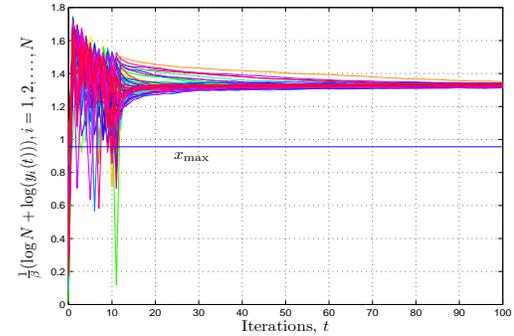}
}
\caption{Simulation Results for Max Consensus.}
\label{Fig2} 
\end{figure}

\begin{figure}
\centering
	\includegraphics[height=4.85cm,width=7.45cm]{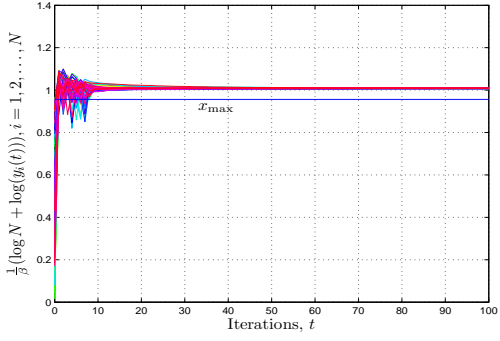}
	\caption{Entries of the consensus soft max result versus iterations $t$, $\beta \!=\! 30$, $\omega \!=\! 10^{-11}$, $h(x) \!=\! \sqrt{\gamma} \tanh(\omega x)$, $\alpha(t)\! =\! \frac{5.03 \times 10^{10}}{t+1}$, $\ao \!\approx\! 5.03 \times 10^{10}$.}
	\label{Fig26} 
\end{figure}

\subsection{Performance of Max Consensus}

In the max consensus simulations, the initial measurements $\{ x_i\}$ are chosen to be uniformly distributed over $(0, 1)$ in Figure~\ref{Fig02} and~\ref{Fig2}. Gaussian noise with zero mean and unit variance is added to receiver nodes. The graph of the sensor network is fixed for all simulations with $N = 75$ sensors.
as shown in Figure~\ref{Fig01}. 
In Figure~\ref{Fig02}, the traditional max consensus algorithm is used and each node always keeps the largest measurement from its neighbors. We can see from Figure~\ref{Fig02} that in the presence of noise, the states of nodes will diverge.

In Figure~\ref{Fig2}, $\hatx_{\text{max}_i}(t^{*})$ from equation \eqref{eq:xmax_estimate} for all nodes are plotted to illustrate the convergence of the soft-max result for different $\beta$ and $a$ values. Note that the actual maximum value is 0.9561 in the simulations. In each of Figures~\ref{Fig2}, $h(x)$ = $\sqrt{\gamma} \tanh(\omega x)$, with $\omega = 0.015$ and $\gamma = 7.5{\rm dB}$, note that $\gamma$ controls the peak transmit power and $\omega$ controls the shape of $h(x)$; $\beta$ is 5 in Figure~\ref{Fig22} and 7 in Figure~\ref{Fig24}. $a$ is chosen as $(N+1)/[2N \lambda_2(\La)h^{\prime}(\bar{y})] \approx \ao$ and the following observations can be made by comparing the two figures: (i) As $\beta$ increases, the estimates of the soft-max of $\x$ are closer to the actual value of the maximum value of $\x$ and (ii) As $\beta$ increases, the convergence is slower, which matches the result in equation \eqref{eq:asymp_covariance_value}. In Figure~\ref{Fig26}, an accurate max estimate is obtained by set $\beta = 30$. It is shown that by properly choosing the non-linear function $h(\cdot)$ and step size $a$, an accurate max consensus can be attained within a few iterations. From Figure~\ref{Fig26}, we can see that the error between the convergence result and the true max is around $0.06$, therefore, Figure~\ref{Fig26} can be a recommended solution for max estimation in sensor networks.

\subsection{Performance of Max Consensus with Shifted Non-linear Bounded Function}

\begin{figure}
\centering
\subfigure[Entries of the consensus soft max result versus iterations $t$, $\beta$ = 7, $\omega$ = 0.015, $N$ = 75, $h(x)$ = $\sqrt{\gamma} \tanh(\omega x)$,  $\alpha(t)$ = 12/($t$+1).] 
{
    \label{Fig_shift_1}
    \includegraphics[height=4.85cm,width=7.45cm]{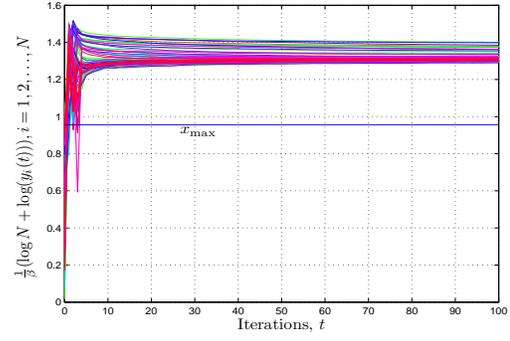}
}

\centering
\subfigure[Entries of the consensus soft max result versus iterations $t$, $\beta$ = 7, $\omega$ = 0.015, $N$ = 75, $h(x)$ = $\sqrt{\gamma} \tanh(\omega (x - T))$, $T = 138.1045$, $\alpha(t)$ = 12/($t$+1).] 
{
    \label{Fig_shift_2}
    \includegraphics[height=4.85cm,width=7.45cm]{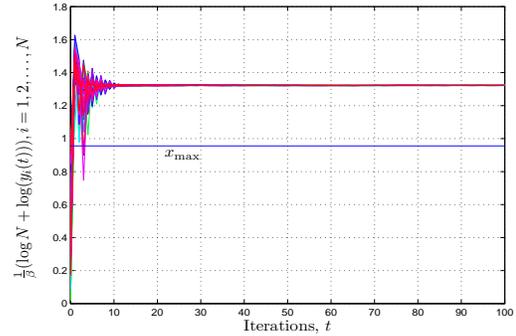}
}
\caption{Simulation Results for Max Consensus Using Shifted Non-linear Bounded Functions.}
\label{Fig_max_consensus_shifted} 
\end{figure}

In the simulation of max consensus using a shifted non-linear bounded function, the initial measurements $\{ x_i\}$ are chosen to be uniformly distributed over $(0, 1)$. Gaussian noise with zero mean and unit variance is added to the receiver nodes, ${\gamma} = 7.5{\rm dB}$ in all simulations. The graph is the same as the max consensus simulation with $N = 75$ sensors. In Figure~\ref{Fig_max_consensus_shifted}, $\hat{x}_{\text{max}_{i}} (t^{*})$ for all the nodes are plotted. In each of Figures, $a = 12$, $\omega = 0.015$ and $\beta$ is $7$. In Figure~\ref{Fig_shift_2}, shifted non-linear bounded functions are used in transmission, and $T$ is chosen to be the sample mean of the mapped initial measurements. By comparing Figure~\ref{Fig_shift_1} and~\ref{Fig_shift_2}, it is shown that using the shifted non-linear bounded function can improve the convergence speed.

\section{Conclusions} \label{sec: conclusions}

A practical approach for reliable computation of the maximum value of local measurements over autonomous sensor networks with no fusion center is proposed. 
The trade-off between estimation accuracy and convergence time is quantified.
It is proved that the sensor network will reach consensus, that is, the state values converge to a random variable whose expectation is the sample mean of the mapped function, and the soft-max can be calculated using the consensus result. The shifted non-linear function used to adjust the transmit nonlinearity is also introduced to make the convergence speed faster. 
The results provide guidelines towards nonlinear transmission design, and algorithm parameter settings to trade-off between estimation error and faster convergence.

\label{eq: xit_Pr_upperBound_step6}

\appendix
\label{appendix: prof_covariance_result}

The Convergence will be slow when $\| \CS \|$ is large, where $\| \CS \|$ is the max eigenvalue of $\CS$. The problem can be formulated as,
\small
\begin{equation}
\| \CS \| = \label{eq: asy_C_formulated_1_cal_1}
\max_{\{x| x\in \mathbb{R}^{N} \; \| \mathbf{x} \| \leq 1 \}} \mathbf{x}^{\rm T} \CS \mathbf{x}.
\end{equation}
\normalsize

Let $\mathbf{U} = [\frac{\mathbf{1}}{\sqrt{N}} \; \mathbf{\Phi}]$, the columns of $\mathbf{U}$ are the eigenvectors of $\La$. Since $\La$ is an Hermitian matrix, the columns of $\mathbf{U}$ form an orthonormal basis of $\mathbb{R}^{N}$. Let $\mathbf{x} = \mathbf{U} \mathbf{z}$ with $\| \mathbf{z}\| \leq 1$, we have
\small
\begin{align}
\mathbf{x}^{\rm T} \CS \mathbf{x} & =  (\mathbf{U} \mathbf{z})^{\rm T} \CS (\mathbf{U} \mathbf{z})   \\
& = \mathbf{z}^{\rm T} \left( \frac{a^2 \sigma^{2}_{n}\mathbf{U}^{\rm T} \mathbf{1} \mathbf{1}^{\rm T} \mathbf{U}}{N} + \frac{\mathbf{U}^{\rm T} \mathbf{\Phi} \mathbf{S}^{\theta_0} \mathbf{\Phi}^{\rm T}\mathbf{U} }{N} \right) \mathbf{z}  \\
& = \mathbf{z}^{\rm T} \left\lbrace   \mathbf{A}_1 + \mathbf{A}_2  \right\rbrace \mathbf{z}^{\rm T}  = \mathbf{z}^{\rm T}  \mathbf{A}_3  \mathbf{z}^{\rm T}, 
\label{eq:pf_norm_AsyCov_matrix_step_2}
\end{align}
\normalsize
where $\mathbf{A}_1 = {\rm diag}\left[a^2 \sigma^2_n, \;\; 0, \; \cdots \;, 0\right]_{N \times N}$, $\mathbf{A}_2 = {\rm diag} \left[0, \; \frac{1}{N} \mathbf{S}_{1,1}, \; \cdots \;, \frac{1}{N} \mathbf{S}_{n-1,n-1} \right]_{N \times N}$ and $\mathbf{S}_{i, i} = \frac{a^2 \sigma^{2}_n}{2a h^{\prime}(\theta_0)\lambda_{i+1}(\La) - 1}, \; i = 1, 2, \cdots, N-1$. $\mathbf{A}_3 = {\rm diag} \left[a^2 \sigma^2_v, \; \frac{1}{N} \mathbf{S}_{1,1}, \; \cdots \;, \frac{1}{N} \mathbf{S}_{n-1,n-1} \right]_{N \times N}$. Equality in \eqref{eq:pf_norm_AsyCov_matrix_step_2} holds since the columns of $\mathbf{\Phi}$ are orthogonal to $\mathbf{1}$ and $\mathbf{S}^{\theta_0}$ is a diagonal matrix which can be calculated as,
\small
\begin{align}
&\mathbf{S}^{\theta_0} = a^2 \int_{0}^{\infty} e^{[ah^{\prime}(\theta_0)\mathbf{B} + \mathbf{I}/2]t} \mathbf{C} e^{\left[ah^{\prime}(\theta_0)\mathbf{B} + \mathbf{I}/2\right]t} \; dt \\
& = a^2 \sigma^{2}_{n} \int_{0}^{\infty} e^{\mathbf{H} t} dt \\
\label{eq: S_matrix_calculated_last}
& = {\rm diag}\left[ \frac{a^2 \sigma^{2}_n}{2a h^{\prime}(\theta_0)\lambda_2(\La) - 1}, \cdots, \frac{a^2 \sigma^{2}_n}{2a h^{\prime}(\theta_0)\lambda_N(\La) - 1} \right], 
\end{align}
\normalsize
where $\mathbf{H}$ is an $(N-1) \times (N-1)$ diagonal matrix and $\mathbf{H}_{i,i} = 2 a h^{\prime}(\theta_0) \lambda_{i+1}(\La) - 1$. Note that \eqref{eq: S_matrix_calculated_last} holds under the assumption that $2a h^{\prime}(\theta_0)\lambda_i(\La) - 1 > 0$ for all $i$, which is same as the requirement in Theorem 5 in \cite{nonLinear2012Sivaraman} that $[a h^{\prime}(\theta_0) \mathbf{B} + \mathbf{I}/2]$ is stable.

Since $\lambda_2(\La)$ is the smallest non-zero eigenvalue, we have
\small
\begin{equation}
\label{eq: compare_lamda_value_2_max}
\frac{1}{N}\frac{ a^2 \NoiseSigma^2}{2a h^{\prime}(\theta_{0})\lambda_{2}(
\La) - 1} \geq \frac{1}{N}\frac{ a^2 \NoiseSigma^2}{2a h^{\prime}(\theta_{0})\lambda_{i}(
\La) - 1}, \; \text{for}\; i>2.
\end{equation}
\normalsize
Therefore, from equation \eqref{eq:pf_norm_AsyCov_matrix_step_2}, \eqref{eq: S_matrix_calculated_last} and \eqref{eq: compare_lamda_value_2_max}, we get,
\small
\begin{align}
\label{eq: proof_done_asy_COV}
\|\CS \| &= \max_{\{x| x\in \mathbb{R}^{N} \; \| \mathbf{x} \| \leq 1 \}} \mathbf{x}^{\rm T} \CS \mathbf{x} \notag \\
&= \max \left\lbrace a^2 \NoiseSigma^2, \; \frac{1}{N}\frac{ a^2 \NoiseSigma^2}{2a h^{\prime}(\theta_{0})\lambda_{2}(\La) - 1} \right\rbrace.
\end{align}
\normalsize

In the following, the optimal $\|\CS^{*} \|$ is calculated together with the corresponding optimal $a = \ao$.
\small
\begin{align}
\label{eq: begin_solve_optimal_COV}
&\|\CS^{*} \| = \min_{\{ a|2ah^{\prime}(\theta_0)\lambda_2(\La) >1 \}} \; \max_{\{x| x\in \mathbb{R}^{N} \; \| \mathbf{x} \| \leq 1 \}} \mathbf{x}^{\rm T} \CS \mathbf{x} \notag \\
& \!=\! \min_{\{ a|2ah^{\prime}(\theta_0)\lambda_2(\La) >1 \}} \max \left\lbrace\! a^2 \NoiseSigma^2, \frac{1}{N}\frac{ a^2 \NoiseSigma^2}{2a h^{\prime}(\theta_{0})\lambda_{2}(\La) \!-\! 1} \!\right\rbrace.
\end{align}
\normalsize

We noticed that the first term $a^2 \NoiseSigma^2$ in equation \eqref{eq: begin_solve_optimal_COV} is a monotonic increasing function of $a$. The monotonicity for the second term can be check by taking the derivative respect to $a$, it is easy to check that the term is decreasing if $\frac{1}{2 h^{\prime}(\theta_{0}) \lambda_{2}(\La)} < a \leq \frac{1}{h^{\prime}(\theta_{0}) \lambda_{2}(\La)}$, and the term is increasing if $a > \frac{1}{h^{\prime}(\theta_{0}) \lambda_{2}(\La)}$.

By checking the value of $\|\CS^{*} \|$ for marginal $a$, we find that the problem in equation \eqref{eq: begin_solve_optimal_COV} is solved by letting,
\small
\begin{equation}
\label{eq: solve_optimal_COV_letEqual}
a^2 \NoiseSigma^2 = \left(\frac{1}{N} \right) \left( \frac{ a^2 \NoiseSigma^2}{2a h^{\prime}(\theta_{0})\lambda_{2}(
\La) - 1} \right).
\end{equation}
\normalsize
By solving equation \eqref{eq: solve_optimal_COV_letEqual}, we get,
\small
\begin{equation}
\label{eq: optimal_set_size_a_calculation_result}
a = \ao = \left( \frac{N+1}{2N} \right) \left( \frac{1}{\lambda_2(\La)h^{\prime}(\theta_0)} \right).
\end{equation}
\normalsize
It is easy to check that $\frac{1}{2 h^{\prime}(\theta_{0}) \lambda_{2}(\La)} <  \ao \leq \frac{1}{h^{\prime}(\theta_{0}) \lambda_{2}(\La)}$. Plug the optimal $\ao$ into the expression of $\| \CS \|$, the corresponding optimal value, $\| \CS^{*} \|$ is given by,
\small
\begin{equation}
\label{eq:asymp_covariance_value_2}
 \| \CS^{*} \| = \left (\frac{N+1}{2 N} \right)^2 \left (\frac{\sigma^2_{n} }{\lambda^2_2(\La)} \right )  \left ( \frac{1}{h^{'}(\theta_0)} \right )^2.
\end{equation}
\normalsize

\bibliographystyle{IEEEtran}
\bibliography{pm_bib_file}
%
%


\end{document}